\documentclass[]{hdsr}

\usepackage{slashed}
\usepackage{tikz}
\usepackage{feynmp-auto}

\begin{document}

\newgeometry{bottom=1.5in}

\volumeheader{0}{0}{00.000}

\begin{center}

  \title{Modern Machine Learning and Particle Physics}
  \maketitle

  \thispagestyle{empty}
  
  \vspace*{.2in}

  \begin{tabular}{cc}
   Matthew D. Schwartz\upstairs{\affilone}
   \\[0.25ex]
   {\small  Department of Physics, Harvard University} \\
   {\small The NSF AI Institute for Artificial Intelligence and Fundamental Interactions}

  \end{tabular}
  
  \emails{
    \upstairs{\affilone}schwartz@g.harvard.edu 
    }
  \vspace*{0.4in}

\begin{abstract}
Over the past five years, modern machine learning has been quietly revolutionizing particle physics. Old methodology is being outdated and entirely new ways of thinking about data are becoming commonplace. This article will review some aspects of the natural synergy between modern machine learning and particle physics, focusing on applications at the Large Hadron Collider. A sampling of examples is given, from signal/background discrimination tasks using supervised learning to direct data-driven approaches. Some comments on persistent challenges and possible future directions for the field are included at the end.  
\end{abstract}
\end{center}


\copyrightnotice

\section{Introduction}
Particle physics is the study of the subatomic constituents of matter: How many are there? What are their properties? How do they interact? There are two basic approaches to answering these questions: a theoretical one and an experimental one. On the theoretical side, we can ask: what possible subatomic particles could there be? Remarkably, there are  constraints due to theoretical consistency of the underlying theory that strongly limit the types of particles possible. For example, there is a direct logical path from the requirement that things do not start appearing out of nowhere (``unitarity'') to the Pauli exclusion principle, which keeps matter from imploding. To the dismay of many theorists, however, there seem to be many more self-consistent theories than the one describing nature, and so experiments are essential. 
The state-of-the art particle experiment is the Large Hadron Collider (LHC) at CERN on the border between France and Switzerland. 
Its major success so far is finding the Higgs boson in 2012. The LHC collides protons together at close to the speed of light. This energy is then converted into mass via $E=mc^2$ thereby forming new particles.
Usually these particles last for only fractions of a second (the lifetime of the Higgs boson is $10^{-22}$ s); hence, the art of modern experimental particle physics involves finding indications that a particle was made even though we never actually see it. The experimental challenge is complicated by the fact that particles of interest are usually quite rare and look nearly identical to much more common backgrounds. For example, only one in every billion proton collisions at LHC produces a Higgs boson, and only one in ten thousand of these is easy to see. Finding new particles in modern experiments is like finding a particular piece of hay in a haystack.
Luckily, hay-in-a-haystack problems are exactly what modern machine learning excels at solving. 

There are two aspects of particle physics that make it unique, or at least highly atypical, as compared to other fields where machine learning is applied. First, particle physics is governed by quantum mechanics. Of course, everything is governed by quantum mechanics, but in particle physics the inherent uncertainty of the quantum mechanical world affects the nature of the truth we might hope to learn. Just like how 
Schr\"odinger's cat can be alive and dead at the same time, a collision at the LHC can both produce a Higgs boson and not produce a Higgs boson at the same time. In fact, there is quantum mechanical interference between the signal process (protons collide and a Higgs boson is produced) and a background process (protons collide without producing a Higgs). The question ``Was there a Higgs boson in this event?'' is unanswerable. To be a little more precise, for a given number of particles $n$ produced, the probability distribution for signal and background, differential in the momenta of the particles produced (the phase space) has the form
\begin{equation}
    d P_{\text{data}}^n = |{\mathcal M}_S +{\mathcal M}_B|^2 dp_1 \cdots dp_n \label{Pdata1}
\end{equation}
Here, ${\mathcal M}_S$ and ${\mathcal M}_B$ are the quantum-mechanical amplitudes ($S$-matrix elements, which are complex numbers) for producing signal and background, and the cross term ${\mathcal M}_S {\mathcal M}_B^\star + {\mathcal M}_B {\mathcal M}_S^\star$ represents the interference. This interference term can be positive (constructive interference) or negative (destructive interference).
Although an individual event cannot be assigned a truth label, the probability of finding a certain set of particles showing up in the detector depends on whether the Higgs boson exists:
finding the Higgs boson amounts to excluding the background-only hypothesis (Eq.~\eqref{Pdata1} with ${\mathcal M}_S=0$). 

In practice, the probability distribution of signal is often strongly peaked, due to a resonance for example, in some small region of phase space. In such regions, background can often be neglected: ${\mathcal M}_S +{\mathcal M}_B\approx {\mathcal M}_S$. In complementary regions, signal can often be neglected: ${\mathcal M}_S +{\mathcal M}_B\approx {\mathcal M}_B$. Thus it is commonplace to approximate the full probability distribution with a mixture model. If we sum over possible numbers $n$ of particles and integrate over the momenta in the observable region, we can then write
\begin{equation}
P_{\text{data}} = \alpha_S \,  P_{\text{S}} + \alpha_B \,  P_{\text{B}}
\end{equation}
with $\alpha_S + \alpha_B=1$. 
That is, we treat the probability distribution of the  data as a linear combination of the probability distributions for signal and background. The goal is then to determine the coefficients $\alpha_S$ and $\alpha_B$, or often more succinctly, whether $\alpha_S$ is non-zero. 
Each measured event gives us a number of particles $n$ and point in $n$-particle phase space $\{p_i\}$, with some uncertainty or binning $dp_1\cdots d p_n$,
 drawn from the true probability distribution $d P^n_{\text{data}}$.
 Only after many draws can we hope to constrain $\alpha_S$. Even within the mixture model approximation, there is still no truth label for individual events. 
  This is different from, say, distinguishing cats from dogs (or alive cats from dead cats) in an image database. For cats and dogs, even if the distributions overlap, there is a correct answer ($\alpha_S=1$ or $0$ for each event). For particle physics, where the distributions overlap, a particle is both signal {\it and} background.

The second way in which particle physics differs from typical machine learning applications is that particle physics has remarkably accurate simulation tools for producing synthetic data for training. These tools have been developed by experts over more than 40 years. Together, they describe the evolution of a particle collision over 20 orders of magnitude in length. The smallest scale the LHC probes is around $10^{-18}$ m, one thousandth the size of a proton. Here the physics is described by perturbative quantum field theory;  particles interact rather weakly and first-principles calculations are accurate. The Higgs boson has a size (Compton wavelength) of $10^{-17}$ m, so it is only at these small distances that we have any hope of examining it. Between $10^{-18}$ m and $10^{-15}$ m, a semi-classical Markov model is used to turn a handful of primordial particles into hundreds of quarks and gluons. Between $10^{-15}$ m and $10^{-6}$ m, the quarks and gluons turn into a zoo of metastable subatomic particles that subsequently decay into hundreds of ``stable'' particles: pions, protons, neutrons, electrons and photons. These then start interacting with detector components and propagating through the material, as described by other excellent parameterized models. The detector model is accurate from $10^{-6}$ m to the $100$ m size of the LHC detectors (the ATLAS detector at the LHC is 46 meters long). The result is a progression from an order-10 dimensional phase space at the shortest distances, to an order-$10^3$ dimensional phase space at intermediate scales, to an order-$10^8$ dimensional phase space of electronic detector readouts channels. Combined, these simulation tools give a phenomenally robust (but embarrassingly sparse) sampling from this hundred million dimensional space. Around one trillion events have been recorded at the LHC, and a comparable number of events have been simulated, providing hundreds of petabytes of actual and synthetic data to analyze. The first stage of the simulations, up to the stable particle level (the $10^3$ dimensional space) is relatively fast:  one million events can be generated in an hour or so on a laptop.
The second stage of the simulation, through the detector, is much slower, taking seconds or even minutes per event. Conveniently, for many applications, the first stage of the simulation is sufficient. 

No human being, and as yet no machine, can visualize a hundred million dimensional distribution. So the typical analysis pipeline is to take all of the low-level outputs and aggregate them into a single composite feature, such as the total energy of the particles in some region. Ideally, a histogram of this feature would exhibit a resonance peak or some other salient indication of signal. We additionally want this feature to have a simple physical interpretation, so that we can cross-check the distribution against our intuition. For the Higgs boson, the ``golden discovery channel'' was two muons (or electrons) and two antimuons (or positrons). A Feynman diagram describing this process looks like
\begin{equation}
 {
\parbox{10mm} {
\includegraphics[width=0.6\columnwidth,trim = {100 0 00 0}]{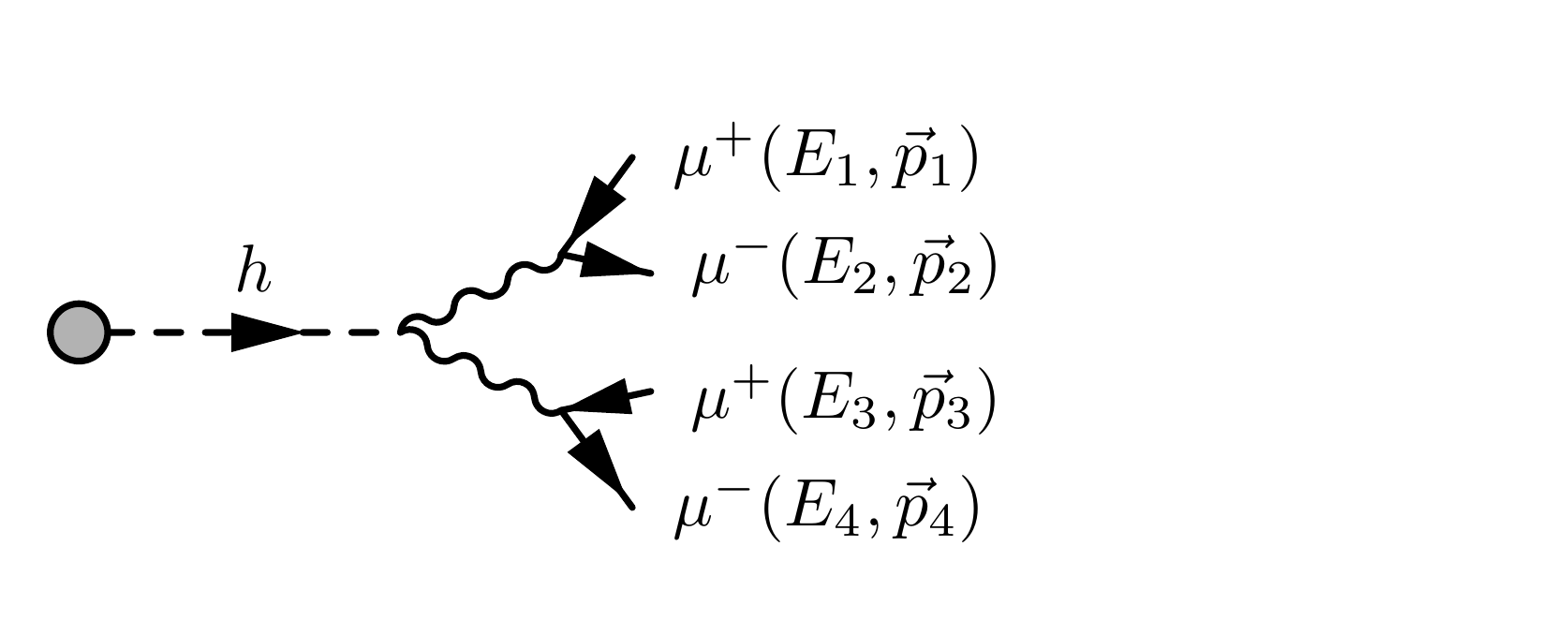}
}
}
\end{equation}
The invariant mass $m=\sqrt{(E_1+E_2+E_3+E_4)^2 - (\vec{p}_1+\vec{p}_2+\vec{p}_3+\vec{p}_4)^2}$, with $E_i$ the energies and $\vec{p}_i$ the momenta,
of the four observed particles is a powerful way to discover the signal.
For the Higgs boson signal, the probability density of this feature has a peak at the Higgs boson mass of 125 GeV where the  background is very small. Unfortunately, only one in every $10^{13}$ proton collisions will give such a signal. If we do not demand that our signal be background-free, and we also do not demand having any physical interpretation such as we have for a feature like mass, 
then we can ask: what feature is the {\it optimal} way of statistically discriminating a signal from its background? Such questions, when supplemented with the enormous amounts of easily produced synthetic data, are ideally suited to modern machine learning methodology.

\section{Supervised learning}

Machine learning (ML) has played a role in particle physics for decades. An emblematic use case is in ``$b$-tagging'': determining whether a given set of particles is associated with a primordial bottom quark. Bottom quarks are around four times heavier than a proton and have properties that help distinguish them from other particles. For example, they tend to travel around half a millimeter away from the collision point before decaying. Technically, the $b$ quark binds with other quarks into metastable hadrons, like the $B_d$ meson, which then decay into particles like muons and pions. One cannot directly measure the distance the particles travel, but by measuring things like the number of decay products, distances among charged tracks, whether there was a muon in the decay, etc., one can accumulate a number of highly correlated features that can be combined to estimate
the probability that there was a $b$ quark involved. Traditionally, the various features might be fed into a shallow neural network or a boosted decision tree to determine a $b$-tagging probability.

$b$-tagging is characteristic of how ML has traditionally (and very successfully) been used in particle physics: physically motivated classifiers are first understood individually and then combined using a relatively simple multivariate technique.
Over the last several years, this paradigm has been replaced by what I like to call {\it modern} machine learning. The modern approach is to feed raw, minimally-processed data, rather than high-level physically-motivated variables, into a deep neural network. The network is then free to find what it thinks is most valuable in the data. For example, with $b$ tagging, a modern machine learning approach is to put all the measured tracks into a recurrent neural network. The network is then trained using labeled simulated data to distinguish signal events ($b$ quarks) and background events (other quarks). This is in contrast to the traditional approach, where the tracks are connected with curves and distilled down to an impact parameter. While the traditional approach works very well, it might for example obtain a factor of 1000 rejection of background quarks while keeping 50\% of the $b$-quarks, the modern approach works much better, rejecting as many as 2000 of the background quarks at the same signal efficiency~\citep{ATLASbtag}. A factor of two in rejection is significant, and impressive evidence that modern machine learning methods are here to stay.

As a second example, consider the problem of pileup mitigation. To understand pileup, it is important to first understand the way modern particle colliders work. At the LHC for example, in order to collide a billion protons per second, the particles are collected into bunches of around $10^{11}$ protons each, with around 3000 bunches circulating in the LHC tunnel at any given time. At these operating parameters, 100 or more protons may collide each time the bunches pass through each other. Of these 100 collisions, only rarely is one a direct hit, i.e., has quarks within each proton strike each other with enough energy to produce something of interest, like a Higgs boson (only one in a billion collisions produce a Higgs boson). When there is a direct hit, often called a primary collision, there are other protons colliding too, called secondary collisions. The protons involved in the secondary collisions disintegrate into essentially sprays of relatively low-energy pions that permeate the detectors. This uninteresting detritus is called pileup. Pileup makes it difficult to ascertain the exact energy involved in the primary collision and contaminates nearly every measurement at the LHC.

There are a number of traditional approaches to pileup removal. One popular method called area subtraction \citep{Cacciari:2008gn} exploits the fact that pileup comprises mostly low-energy particles that are nearly isotropically distributed in pseudorapidity (pseudorapidity measures how close to a beam a particle is) and azimuthal angle. Area subtraction recalibrates the event based on the amount energy deposited in some region of the detector where products from the primary collision are believed to be absent. Another method, used extensively by the CMS collaboration, is called charged hadron subtraction~\citep{CMS:2014ata}. This method uses the fact that charged particles leave tracks, so that one can match the tracks to either the primary collision or a secondary collision. The ones that come from the secondary collision are then removed from the event. Both of these methods are effective but rather coarse: area-subtraction  works only on average, and charge hadron subtraction cannot account for the neutral particles. Neither method attempts to locate all the pileup radiation in each individual event.

\begin{figure}[t]
\hspace{-2cm}
\includegraphics[]{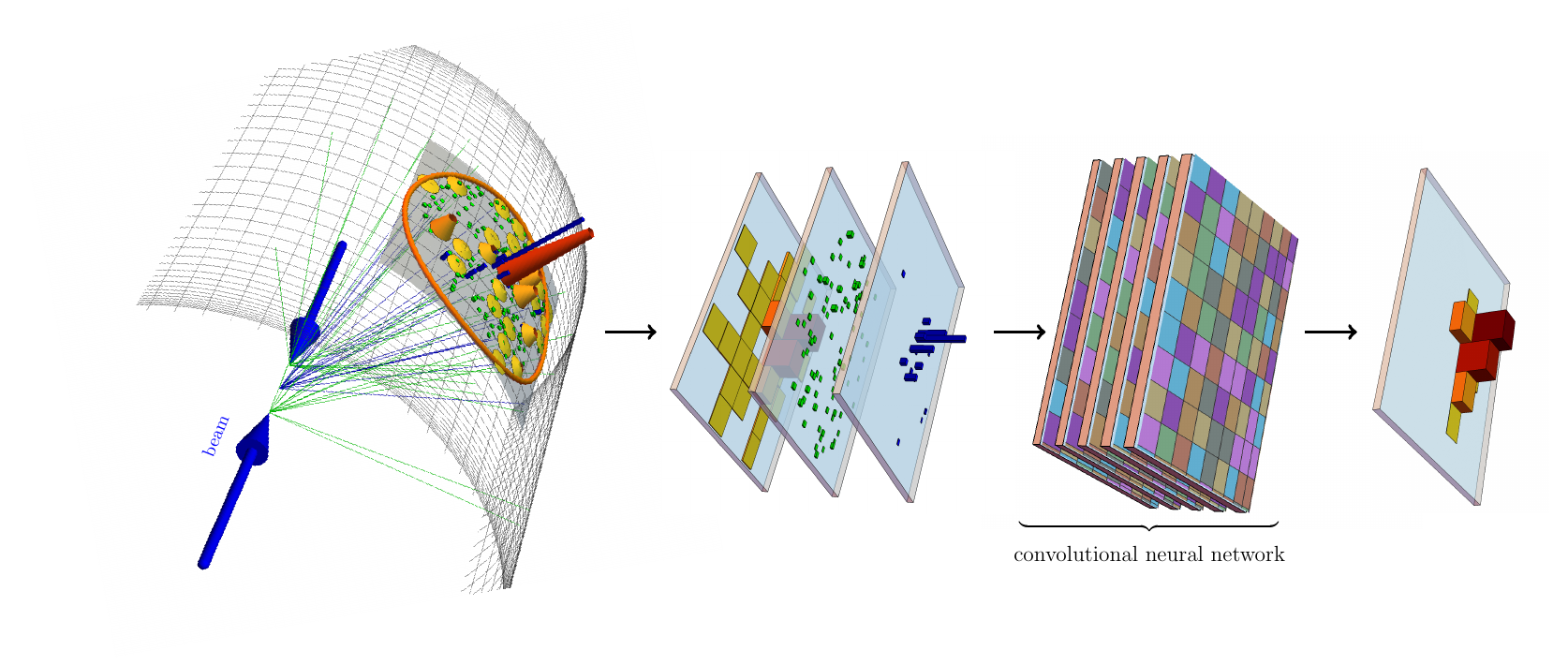}
\vspace{-1cm}
\caption{An illustration of the deep convolutional neural network pipeline used in~\citep{Komiske:2017ubm} for pileup removal. Collider data is categorized
and input to the network, which regresses the particle distributions of the pileup-free event.}
\label{fig:netarch}
\end{figure}

A modern machine learning approach to pileup removal is the PUMML (PileUp Mitigation with Machine Learning) algorithm~\citep{Komiske:2017ubm}. PUMML builds upon the idea of a event image: the energy deposited into a given region of the detector is translated into the intensity of a pixel in an image~\citep{Cogan:2014oua}. For PUMML, three images are constructed: one for charged particles from the primary collision point, one for charged particles from the secondary collisions points and a third from the neutral particles. These three images are fed into a convolutional neural network (CNN) which attempts to regress out a fourth image showing neutral particles from the primary interaction only. The algorithm can be trained on synthetic data, where truth information about the origin of the neutral particles is known, and then applied to real data where the truth is not known. A sketch of the algorithm is shown in Fig.~\ref{fig:netarch}. The PUMML algorithm is extraordinarily effective: it succeeds in reconstructing the full distribution of particles from the primary collision on an event-by-event basis independently of the number of synchronous secondary collisions. Although it foregoes some physics knowledge (like the isotropy of pileup radiation exploited by area subtraction), this modern machine learning approach reaps enormous gains in efficacy.

We have seen how a recurrent neutral architecture, originally developed for natural language processing, was useful in $b$-tagging. We have also seen how a convolutional neural network, developed for image recognition, was useful for pileup removal. In a sense, a large part of what has been done so far in supervised learning in particle physics can be characterized as a series of similar exercises: a ML technique developed for an entirely different purpose is adapted for a particle physics application. A fair comparison of a variety of these approaches was recently made for the problem of boosted top-tagging. The top quark is the heaviest known quark. When it is produced with energy much in excess of its rest mass, as it commonly is at the LHC, it will decay to a collimated beam of particles, a ``jet'', that is difficult to distinguish from a collimated beam of particles {\it not} coming from a top-quark decay. Indeed, there can be 10,000 times more of these background jets then there are top jets. 
The traditional approach to distinguishing top jets from background jets focuses on physically-motivated distinctions: the top quark has a mass, the top jet usually has three subjets, corresponding to the three light quarks into which a top quark decays, etc.~\citep{Kaplan:2008ie}. The modern machine learning approach is to throw the kitchen sink into a neural network and hope it works.

\begin{figure}
    \centering
\begin{tikzpicture}
  \node at (0,0)  {\includegraphics[width=0.9\columnwidth]{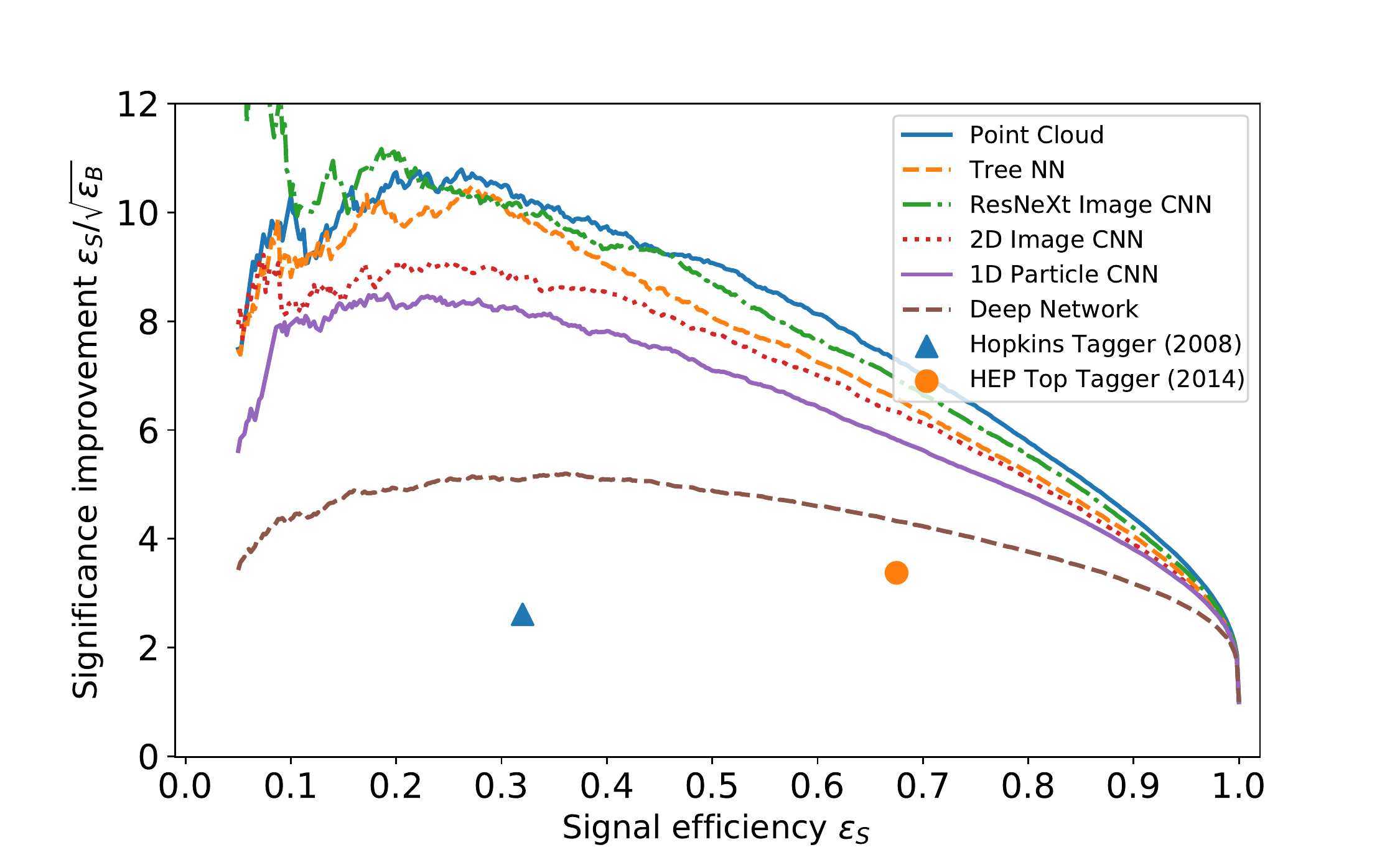}};
\end{tikzpicture}
    \caption{Performance of modern machine learning methods for the task of identifying top quarks. The vertical axis is $\frac{\epsilon_S}{\sqrt{\epsilon_B}}$: the signal (top quark) efficiency divided by the square-root of background (non-top processes) efficiency. This roughly corresponds the number of standard deviations of improvement achievable in a search using the corresponding discriminant (compared to just counting all the events in the samples). 
     Modern machine-learning approaches (curves) do significantly better than traditional approaches (points).
   Figure adapted from~\citep{Kasieczka:2019dbj}.}
    \label{fig:tops}
\end{figure}

Fig.~\ref{fig:tops} shows a comparison between traditional approaches (points) and modern machine learning approaches (curves). At the time, the traditional approach was a game-changing advance in particle physics. Before that, people did not even hope to find tops at these energies. Nevertheless, we can see from this figure that the modern machine learning algorithms noticeably outperform traditional physics-motivated algorithms.
In more detail, the curve labeled ``Deep Network'', modeled after ~\citep{ATLAStoptag},  takes as input the momenta components of the particles (up to 600 inputs) and passes them through a four-layer densely-connected feed-forward neural network. The ``1D Particle CNN'' network uses similar inputs but pipes them through a 1-dimensional convolutional network. This more sophisticated network architecture leads to noticeable improvement. The ``2D Image CNN'' curve uses an image-based convolutional network, where the input image pixel intensity is proportional to the the energy deposited in a region of the detector~\citep{Kasieczka_2017}, as in the pileup example discussed earlier. The ``ResNeXt Image CNN'' curve uses the much more sophisticated ResNeXt convolutional network architecture~\citep{xie2016aggregated}. While there is performance gain, it is at the expense of 1.5 million trainable network parameters (in contrast to around 500,000 parameters for the CNN). The ResNeXt curve is unstable at small signal efficiency due to statistical limitations of the testing samples.  
The ``Tree NN'' curve uses a little more physics input: it organizes the particles' momenta into a binary tree similar to the semi-classical branching picture of sequential particle decays. With only 34,000 parameters, this network performs comparably to the highly-engineered ResNeXt one, giving hope that physical insight may not entirely be disposable. Finally, the curve labeled ``Point Cloud'' uses unordered point-cloud type representation of the inputs called Particle Net~\citep{Qu_2020}; the architecture is based on a Dynamic Graph CNN developed for computer vision~\citep{DGCNN}. More details of all of these algorithms, and some others, can be found in~\citep{Kasieczka:2019dbj}. The superior performance of these neural networks over traditional classifiers makes it clear that modern machine learning methods are finding many more differences between signal and background than were identified by physical reasoning. It remains an open question whether any of these differences can be understood and explained in some simple human-interpretable way. I will return to this question in Section \ref{sec:outlook}.

\section{Data-driven approaches}
All the techniques described above heavily exploit our ability to  generate synthetic data sets for training.
Although the simulations are highly sophisticated, and reproduce the data over 20 orders of magnitude in length scale, they are not engineered to reproduce all of the subtle correlations that modern machine-learning methods might be exploiting. Indeed, until the modern machine learning revolution, there was not a strong motivation to ensure that the correlations were all correct. 
A commonplace, and often implicit, belief is that, although correlations in the synthetic data may not be exactly the same as correlations in real data, the ML methodology should still work. But, until we know for sure, it is hard to assign uncertainties to the output of the ML algorithms on actual data. An alternative to using synthetic data would be to train on real data. Unfortunately, while synthetic data sets have truth labels for training, because we know how we generated them, real data does not. 
Moreover, as mentioned in the introduction, there is no actual ground truth in the real world: in physics, each data point is both signal {\it and} background, to some extent. Even when quantum mechanical interference is small (as it often is), the data is at best mixed signal and background, so it is not immediately clear how to use data directly for training. There are two ways to proceed. First, we can try to train the network directly on the real data despite its impurity. Second, we can use ML to determine how well the simulations agree with the data, and then try to improve the simulations.  Both approaches have already received some attention in particle physics and are currently being explored using LHC data.

An important observation relevant for training directly on data is that although actual data does not come with labels, it is possible to find particularly clean events where labelling can be done unambiguously. For example, top quarks almost always come in pairs (a top and an anti-top). One can restrict to events where, say, the anti-top decays to a muon and a $b$-jet that are cleanly tagged. Then the rest of the event provides a clean top-quark data point. This tag-and-probe method has been a mainstay of experimental particle physics since well before machine learning, and is a useful way to gather truth-labeled samples for calibration.

Another more machine-learning oriented approach is to train directly on mixed samples. For example, one can use a sample of events with one jet and a $Z$ boson, and another sample with two jets. In these samples, it is expected that the fraction of jets coming from a quark is different from the fraction from a gluon (roughly 80\% quark in $Z$+jet and $40\%$ quark in the dijet sample~\citep{Gallicchio:2011xc}). Then one can train the neural network to distinguish these two samples. Such {\it weakly supervised} methods do not try to learn the separate properties of quark and gluons jets, only differences between them. They 
work surprisingly well, either when trained on high-level classifiers like mass~\citep{Metodiev_2017} or when trained with a image CNN~\citep{Komiske:2018oaa}. Such studies foretell a future in which the simulations can be done away with altogether and the data used directly for both training and validation. 

There are a number of fully unsupervised approaches also being developed for applications at the LHC. One example is the JUNIPR framework, which attempts to learn the full differential distribution $\frac{d^n P}{dp_1 \cdots d p_n}$ of the data using machine learning~\citep{Andreassen:2019txo}. JUNIPR has a network architecture scaffolded around a binary jet-clustering tree, similar to the highly effective ``Tree NN'' shown in Fig.~\ref{fig:tops}. Using the tag-and-probe method or weakly supervised learning, one can then train JUNIPR on separate samples to get different probability functions. Doing so lets us go beyond the typical likelihood-free inference approach used in supervised learning applications. For example,
comparing these learned functions can discriminate different samples and find features of interest. Alternatively, a method like JUNIPR can be trained on data and then events can be drawn from the learned probability distributions for data augmentation. Thus, JUNIPR can act like a kind of simulation itself, but with all the elements learned rather than built upon microphysical models. Such methodology could dovetail well with developments in probabilistic programming approaches, as in~\citep{Baydin:2019fap}.

Continuing on the line of improving the simulations, as discussed above these simulations have different components. The short-distance simulation, which produces hundreds of particles using quantum field theory is relatively fast (of order microseconds per event),
while simulating the propagation of these particle through the detector can be significantly slower (of order seconds or minutes per event). Indeed, a significant fraction of all LHC computing time is devoted to running detector simulations. To ameliorate this computing problem, one might turn to an unsupervised learning method like CaloGAN~\citep{Paganini_2018}.
CaloGAN uses a generative adversarial network to mimic the detector simulator. 
With CaloGAN, a first network produces events and a second adversary network tries to tell if those events are from the real detector or the neural network one. Once trained, the NN simulator can be used at a cost of as little as 12 microseconds per event: a five order-of-magnitude speed up compared to the full simulation. Such approaches are extremely appealing, particularly for higher-luminosity future LHC runs where all the computing resources in the world would not be enough to simulate a sufficient number of events.

Rather than learning to reproduce and generate events similar to the particle-level simulation (like JUNIPR) or the detector simulator (like CaloGAN), one can instead learn just the places where the simulation is inaccurate. For example, one could train an unsupervised model on the synthetic data and the real data, and then when the two differ reweight synthetic data to look like real data. A proof-of-principle implementation of this idea is OmniFold~\citep{Andreassen:2019cjw}. OmniFold learns the mapping from simulation to data. Then one can try to invert the mapping to effectively remove the effects of the detector simulation. The process of removing detector effects in particle physics is called unfolding. Unfolding is typically a laborious process, done for each observable separately. OmniFold uses ML methods to learn how the detector affects each event, so that {\it any} observable can be unfolded using the same trained network. This could be a game-changer for experimental analyses, speeding them up by many orders of magnitude.

Finally, it is worth mentioning one more issue that has received some attention in applying ML methods directly to data. A potential problem with ML methods is that they can be so powerful that cutting (refining the event selection) on a learned classifier can sculpt the background to look like the signal. Such sculpting can be misleading if there is no signal actually present, and it can complicate the extraction of signal events from data. To deal with this, one can train the network to learn not the optimal discriminant but an optimum within the class of discriminants that do not sculpt the background in some undesirable way~\citep{Louppe:2016ylz}. Similarly, finding two uncorrelated observables that together optimize a discrimination task can be useful for data-driven sideband background estimation~\citep{Kasieczka:2020pil}. This kind of hybrid approach, where some supervised training is used to guide a data-driven estimation technique, is a very promising area for future development of ML for particle physics.

\section{Outlook \label{sec:outlook}}
In the relatively few years that modern machine learning has existed, it has already made traditional collider physics obsolete. In the past, physicists, including myself, would devote their efforts to understanding signatures of particular particles or processes from first-principles: why should a stream of pions coming from a $W$ boson decay look different than a stream coming from an energetic gluon? Now we simply simulate the events, and let neural network learn to tell the two samples apart. Even a relatively simple dense network with 10 lines of python code can blow the traditional discriminants out of the water. Progress so far has come from taking algorithms masterly engineered for other applications, like convolutional neural networks, and shoehorning the collider data into a format that these algorithms can process. Unfortunately doing so can inhibit the extraction of any kind of physical understanding from the network itself.

Another aspect of particle physics discussed in this review, for which machine learning is also ideally suited, is in data-driven methodology. For example, the weak-supervision paradigm was discussed, where a classifier is trained on data samples that have mixtures of signal and background. Unfortunately, there is no foolproof way to demonstrate that the data-driven approaches are superior to current experimental efforts; this is in contrast to, say, using ResNeXt to distinguish top quarks, where the superiority over traditional theoretical methods can be validated purely on simulation. While tools like OmniFold might make some aspects of experimental analysis obsolete, they also have many precarious failure modes. For example, we often look for signals of new particles with very low production rates. Such signals can be faked by very rare processes, in far tails of probability distributions, like the coincidence of a particle decay and a cosmic ray, or two successive collisions that happen to temporarily overload a particular sensor. While it is easy to train an algorithm to reproduce the majority of a model's output, it is much harder to train it to reproduce every nuance of the model. 

The applications discussed in this review were limited to collider physics. In fact, machine learning is infiltrating all aspects of particle physics, from neutrinos, e.g., ~\citep{Aurisano:2016jvx}, to string theory, e.g. ~\citep{RUEHLE20201}. One very promising area of application is towards improving non-perturbative calculations of the properties of matter in the framework of lattice quantum chromodynamics (QCD). Although QCD is a complete theory, and quantities such as the mass of the proton are in principle calculable with it, actually performing the calculations can be very time consuming and the current state-of-the-art is not as accurate as we would like. Machine learning has the potential to improve the way lattice QCD calculations can be done with learned approximations or by more efficiently sampling the configuration space~\citep{Shanahan:2018vcv}. If ML can be shown to be  scaleable and effective, such techniques could revolutionize our ability to calculate properties of matter, from form factors relevant for dark matter searches to parton distribution functions for nuclear physics experiments. 


We are still in the earliest days of exploring the interface between machine learning and particle physics. An open question that has been receiving increasing amounts of attention in recent years is how we might gain deep physical understanding from the machine learning output. So far, success has amounted mostly to reproducing known interpretations. For example, one can combine an ML classifier and a traditional observable to see if the machine has effectively incorporated all the information that the observable contains. Such exercises can explain why the machine works as well as the traditional observable; the subtle correlations the network leverages to outperform simpler approaches, however, do not have simple interpretations. 

A more general approach to human-interpretable learning might look similar to ``AI Feynman'' \citep{Udrescu:2019mnk}, where the authors used symbolic regression to learn 100 equations from the Feynman Lectures on Physics from noisy numerical sampling of those equations. The domain of functions it could learn were compositions of elementary functions -- polynomials, logarithms, trigonometric functions, etc. There is a sense in which this kind of program can be successful because human-interpretable equations are generally fairly simple; hence, we can delineate a finite domain to draw from. But, to be fair, just because an equation can be written with a few characters, does not make it simple. The Dirac equation $\slashed{\partial} \psi = m \psi$ comprises five symbols, but it takes twenty years of education to understand it. Is $\sin(x)$ a simple function? Not to a 5th grader. Is $x^2$ a simple function? Not to a 3 year old. Perhaps the machine learning output is not understandable to us, no matter how old we are. 

I see three possible ways that the interpretability conundrum may be resolved. First, we may find ways to interpret the machine's output along traditional lines. Maybe the traditional methods are weak because we just missed something. I have no doubt that there are some things we will learn this way.
Second, the understanding we seek may simply be beyond human intelligence. It takes at least 20 years to learn quantum field theory, but maybe it would take more than 100 years, or 1000 years to learn what we need to go beyond quantum mechanics. In this case, we must content ourselves with the knowledge that the machine understands the underlying physics, even if we never will. I accept that there are better athletes than I am, better physicists, better artists, better philosophers, etc. I do not need to understand economics myself to benefit from sound economic policy. The third possibility, which I think is the most promising, is that we will build a new language and new set of tools for interpreting the machine's output. I see no limit to our capacity to compartmentalize, simplify, and extrapolate. For example, consider the function $\ln x$. How do we understand it? We can plot it, we can define it as an infinite series, we can think of it as the inverse of $e^x$, the integral of $\frac{1}{x}$, etc. When we first encounter a logarithmic distribution, it may not even be obvious to give it a name, but after a while we start to see its universality and advantage. When visiting a new culture, or hearing a new type of music, or playing a new game, we naturally go from disorientation to comprehension, often constructing a new implicit or explicit language for the new experiences. For machine learning, the path forward may be similar to the other human endeavors: we need to establish a set of predicates for understanding from which we can build new intuition, rather than trying to force the machines into our traditional worldview. In other words, we may humbly need to learn the machine's language, rather than ask the machines to speak ours. In any case, whether or not we understand the machines, they are here to stay. I feel more optimistic about the possibility of transcendental progress in fundamental physics now than at any other time in my career, even if I myself may not be able to comprehend the final theory.

\label{sec1}

\restoregeometry
\newgeometry{bottom=0.5in}

\subsection*{Acknowledgments}
If not for the persistence of Xiao-Li Meng, this article would not have been written. I would also like to thank Ben Nachman for comments on the manuscript. This work is supported by the US Department of Energy, under grant DE-SC0013607 and the National Science Foundation under Cooperative Agreement PHY-2019786 (The NSF AI Institute for Artificial Intelligence and Fundamental Interactions).






\printbibliography

\end{document}